\documentclass[11pt,a4paper,twocolumn]{article}

\usepackage[utf8]{inputenc}
\usepackage[T1]{fontenc}
\usepackage{amsmath,amssymb,amsfonts,amsthm}
\usepackage{mathtools}
\usepackage{geometry}
\usepackage{booktabs}
\usepackage{array}
\usepackage{graphicx}
\usepackage{xcolor}
\usepackage{hyperref}
\usepackage{abstract}
\usepackage{titlesec}
\usepackage{setspace}
\hypersetup{
	colorlinks=true,
	linkcolor=blue!70!black,
	citecolor=blue!70!black,
	urlcolor=blue!70!black,
	pdftitle={A constrained mixed variational formulation with Legendre-dual defects for incremental irreversible thermodynamics},
	pdfauthor={Mostafa Rostampour},
	pdfsubject={Variational formulation for irreversible thermodynamics}
}

\titleformat{\section}{\large\bfseries}{\Roman{section}.}{0.5em}{}
\titleformat{\subsection}{\normalsize\bfseries}{\Alph{subsection}.}{0.5em}{}

\newtheorem{proposition}{Proposition}

\newtheorem{remark}{Remark}

\title{\textbf{A constrained mixed variational formulation with Legendre-dual defects for incremental irreversible thermodynamics}}

\author{Mostafa Rostampour$^{*}$\\
	\small Department of Physics, Amirkabir University of Technology, Tehran, Iran\\
	\small $^{*}$Corresponding author: \href{mailto:mostafarostampour@aut.ac.ir}{mostafarostampour@aut.ac.ir}; ORCID: \href{https://orcid.org/0009-0000-1988-5660}{0009-0000-1988-5660}.}

\date{}

\begin{document}
	\onehalfspacing
	\twocolumn[
	\begin{@twocolumnfalse}
		\maketitle
		\begin{abstract}
			\noindent
			We present a constrained mixed variational formulation for incremental irreversible thermodynamics. A single augmented action combines a linearized contact-increment penalty, equilibrium and dissipative Fenchel--Young defects, a storage--transport functional, and multipliers enforcing the constitutive increments. Under frozen-coefficient and boundary assumptions, stationarity yields the linearized contact condition, the Tisza relation, the Onsager flux--force law, and the implicit balance equation. An exact-elimination theorem shows that the contact and multiplier fields can be reconstructed algebraically from a solution of the reduced weak balance problem. We also prove off-shell invariance of the equilibrium--storage--multiplier sector under an equilibrium-sector partial Legendre transformation; invariance of the full augmented action is not claimed. For Fourier heat conduction, we derive a weighted drift identity and a per-step energy-drift bound for the energy-representation three-term incremental functional. In one spatial dimension, a small-time-step estimate yields a uniform $H^1$ bound and a discrete $L^2(0,T_f;H^2)$ bound, leading to cumulative $O(\delta t)$ drift over fixed time intervals. A conservative mixed discontinuous Galerkin (DG0)--Raviart--Thomas (RT0) discretization with fully implicit conductivity $\kappa(T)=1+T^2$ preserves discrete energy to roundoff. A separate lumped-mass piecewise-linear benchmark reproduces the predicted finite-step drift.
		\end{abstract}
	\end{@twocolumnfalse}
	
	]
	
	\section{Introduction}
	
	Reconciling the PDEs of irreversible thermodynamics with a genuine
	variational principle has remained a central question in mathematical physics
	since Poincar\'e observed that dissipative evolution equations cannot be
	derived from standard Euler--Lagrange equations~\cite{Poincare1908,Magnenet2020}.
	The inverse problem of the calculus of variations imposes Helmholtz-type
	self-adjointness conditions on the Fr\'echet derivative of a candidate
	Euler--Lagrange operator, subject to standard regularity and domain
	hypotheses~\cite{Santilli1978,Olver1993,Magnenet2020}. Irreversible transport
	equations contain first-order time derivatives that break time-reversal
	symmetry, and for this reason their standard differential operators are not
	self-adjoint~\cite{Magnenet2020}. Classical responses to this difficulty
	include multiplying by an integrating factor, doubling the degrees of freedom
	through an adjoint system, or appending a Rayleigh--Onsager dissipation
	function~\cite{Bateman1931,Biot1955,Kobe1986,Grmela1997,Mielke2011,%
		GayBalmaz2017a,GayBalmaz2017b,Bravetti2019,Magnenet2020}. Incremental
	variational updates have also been used for dissipative solids and
	continua~\cite{OrtizStainier1999}.
	
	More recently, Magnenet and Schmittbuhl~\cite{Magnenet2020} introduced a
	time-discrete incremental variational formulation grounded in Callen's
	axiomatic thermodynamics~\cite{Callen1985}, the nonequilibrium thermodynamics
	of De Groot and Mazur~\cite{DeGroot1984}, and Extended Irreversible
	Thermodynamics (EIT)~\cite{Jou1996}. They represent extensive quantities as
	spatial 3-forms over a material volume element $\Omega = dx \wedge dy \wedge dz$
	and introduce specific cumulated flows $\boldsymbol{\psi}^\alpha$ related to
	dissipative fluxes $\mathbf{j}^\alpha$ along the thermodynamic trajectory by
	$\dot{\boldsymbol{\psi}}^\alpha = \mathbf{j}^\alpha$. This construction yields
	a differential 5-form $\Theta$ that behaves as a dynamical absolute integral
	invariant ($\mathcal{L}_\zeta \Theta = 0$) along the evolution vector field
	$\zeta$ whenever the Tisza stiffness matrix $T_{ab}$ and the Onsager mobility
	matrix $L^{\alpha\beta}$ are symmetric~\cite{Tisza1966,Magnenet2020}.
	Discretizing time into increments $\delta t$ and replacing the antisymmetric
	exterior products by symmetric quadratic forms then gives a convex incremental
	functional. Its minimization reproduces the implicit time-stepped balance PDEs
	and allows efficient reduced-order simulations through the Incremental Ritz
	Method (IRM)~\cite{Magnenet2020}.
	
	Two recent works of Ren~\cite{Ren2026,Ren2026b} develop path-dependent
	entropic Lagrangians with history terms and power-port decompositions for
	irreversible thermodynamics. The action proposed here differs in its
	incremental, time-discrete organization on a pointwise extended contact fibre
	and in its explicit quadratic constitutive defects. We derive a drift identity
	for the energy-representation three-term functional of
	Ref.~\cite{Magnenet2020} and a conservation result for a conservative mixed
	balance formulation. We do not claim that analogous identities cannot be
	formulated within path-dependent approaches; comparing the frameworks on a
	common benchmark, for instance with state-dependent mobility, remains a
	natural next step.
	
	Motivated by the versatility of the framework in Ref.~\cite{Magnenet2020},
	this work explores three complementary refinements that arise in passing from
	the phase-space 5-form $\Theta$ to the reduced quadratic functional. The first
	concerns the \emph{Legendre self-duality of the constitutive defect}. The
	reduced functional relies on the Tisza Hessian
	$T_{ab} = \partial^2 \phi_{\mathrm{eq}}/\partial e^a \partial e^b$, whose
	positive-definite rescaling $g_{ab} = \chi T_{ab}$ is the Weinhold metric in
	the energy representation and the Ruppeiner metric in the entropy
	representation~\cite{Weinhold1975,Ruppeiner1979,Ruppeiner1995}. The two are
	conformally related by $g^W = T\,g^R$~\cite{Salamon1984,Mrugala1991}. The Weinhold and Ruppeiner metrics are not Legendre invariant, whereas
	the GTD metrics are Legendre invariant by construction~\cite{Quevedo2007}.
	
	The companion paper develops a separate Quevedo-type completion of the defect;
	no identification with the standard Quevedo pullback is made here. The second
	concerns the
	\emph{energy--entropy representation change and the spatially varying
		conformal factor}. In Sec.~IV.B of Ref.~\cite{Magnenet2020}, the authors return
	from the entropy to the energy description before deriving their Eq.~(54), and
	the resulting energy increment turns out to be the pairing of $\nabla\ln T_k$
	with the heat-flux increment, as Proposition~\ref{prop:energy_picture} makes
	precise. The third concerns a \emph{constrained mixed variational principle}:
	the reduced storage--transport functional is naturally written in the
	intensive increments, whereas the Tisza and Onsager relations constrain
	independent extensive and flux increments. We therefore introduce auxiliary
	multipliers that impose the constitutive relations while the
	storage--transport sector generates the balance equation.
	
	These elements are combined in a single augmented scalar action. In this
	action, (i) the equilibrium residual is represented by a quadratic defect with
	a proven partial-Legendre duality, (ii) the dissipative residual is represented
	by an algebraically dual flux--force defect that is itself a Fenchel--Young gap
	for a general convex dissipation potential, and (iii) multiplier variations
	impose the Tisza and Onsager residual constraints while the storage--transport
	sector yields the balance equation on the constraint set. The construction is
	complementary to existing variational formulations for irreversible
	thermodynamics. Gay-Balmaz--Yoshimura~\cite{GayBalmaz2017a,GayBalmaz2017b}
	provide a Lagrangian formulation on the full state space, with nonholonomic
	constraints built into the variational principle; Mielke's GENERIC
	framework~\cite{Mielke2011} combines a Poisson bracket with a dissipative
	metric and derives the evolution from an entropy-production principle. The
	present formulation differs in three structural choices: it is incremental and
	time-discrete, it separates the Tisza and Onsager constraints as independent
	Lagrange-multiplier blocks, and it represents each constraint through a
	Fenchel--Young defect whose Legendre duality is established separately. These
	choices make the exact-elimination result of
	Proposition~\ref{prop:exact_elimination} available, at the cost of introducing
	additional multiplier fields. A direct numerical comparison with the above
	frameworks on a common benchmark remains an interesting open question.\newline The resulting structure leads to several concrete results. First, the
	multiplier and contact fields admit an exact algebraic elimination: a
	stationary point of the augmented action is equivalent to a solution of the
	reduced weak balance problem, with all auxiliary fields reconstructed
	algebraically from the reduced solution
	(Proposition~\ref{prop:exact_elimination}). This is verified numerically on the CG1 discretization
	(Remark~\ref{rem:prop3_verification}). Second, the
	equilibrium--storage--multiplier sector is off-shell invariant under the
	equilibrium-sector partial Legendre transformation, with an affine multiplier
	law (Proposition~\ref{prop:eq_storage_invariance}). Third, the full-Hessian
	defect is self-dual under the locally defined full Legendre transformation
	whenever the signed Hessian is positive-definite
	(Proposition~\ref{prop:full_defect}). Fourth, for the energy-representation
	three-term functional of Ref.~\cite{Magnenet2020}, we derive a weighted drift
	identity and a per-step $\mathcal{O}(\delta t^2)$ drift bound
	(Proposition~\ref{prop:drift_bound}); in one spatial dimension this yields a
	small-time-step a priori estimate and a fixed-time cumulative
	$\mathcal{O}(\delta t)$ drift bound
	(Proposition~\ref{prop:1d_weighted_stability}). Finally, a conservative mixed
	$DG_0$--$RT_0$ discretization with $\kappa(T)=1+T^2$ preserves discrete energy
	to roundoff (Proposition~\ref{prop:dg_rt_conservation}). The present
	formulation establishes off-shell duality for the equilibrium sector and exact
	elimination for the linearized mixed system. Extending off-shell Legendre
	invariance to the complete nonlinear augmented action and developing a full
	saddle-point multiplier discretization remain open directions.\newline \section{Extended Contact Fibre and Legendre-Dual Constitutive Defects}
	\label{sec:extended_contact}
	
	\subsection{Kinematics and Continuum Balance Coupling}
	In equilibrium contact thermodynamics~\cite{Hermann1973,Mrugala1978,Mrugala1991,Quevedo2007,Bravetti2019}, a system with $n$ extensive degrees of freedom is modeled on a $(2n+1)$-dimensional contact manifold. For transport on a spatial domain $\Omega \subset \mathbb{R}^d$ with volume element $d\omega$, let $m=n$ denote the number of flux families and let each cumulated flow $\boldsymbol{\psi}^a(x,t)$ and conjugate force $\mathbf{X}_a(x,t)$ be a spatial vector with $d$ independent components. The pointwise extended contact fibre therefore has $n+nd$ independent extensive-type coordinates and the same number of conjugate coordinates, in addition to the local potential density:
	\begin{equation}
		\dim \mathcal{T}_{\mathrm{ext}} = 2n + 2nd + 1.
		\label{eq:extended_dimension}
	\end{equation}
	
	\begin{remark}[Ambient configuration space]
		\label{rem:ambient}
		The relevant functional setting is that of the reduced weak problem, $\delta I\in H^1(\Omega;\mathbb{R}^n)$; the conjugate fields are obtained from the potential density, and no higher regularity of the section is assumed.
	\end{remark}
	
	Its local coordinates may be written as $Z^A = \{\phi, e^a, \psi^{ai}, I_a, X_{ai}\}$, where $a = 1,\ldots,n$ and $i = 1,\ldots,d$. The extended Gibbs contact 1-form is
	\begin{equation}
		\Theta_{\mathrm{ext}} = d\phi - I_a\,de^a - \sum_{a=1}^{n}\sum_{i=1}^{d}X_{ai}\,d\psi^{ai}.
		\label{eq:extended_gibbs}
	\end{equation}
	A local thermodynamic state space $\mathcal{E}_{\mathrm{ext}} \subset \mathcal{T}_{\mathrm{ext}}$ is an $(n+nd)$-dimensional Legendrian submanifold, parametrized by $(e^a,\psi^{ai})$, on which $\varphi^*(\Theta_{\mathrm{ext}}) = 0$ and hence
	\begin{equation}
		d\phi = I_a\,de^a + \sum_{a,i}X_{ai}\,d\psi^{ai},
		\label{eq:legendrian_condition}
	\end{equation}
	with $I_a = \partial\phi/\partial e^a$ and $X_{ai} = \partial\phi/\partial\psi^{ai}$. We use compact vector notation $\mathbf{X}_a \cdot d\boldsymbol{\psi}^a \equiv \sum_i X_{ai}\,d\psi^{ai}$.
	
	For a finite increment along a path $z(\theta)=(e(\theta),\psi(\theta))$ on the Legendrian submanifold, the exact contact relation is the path integral
	\begin{equation}
		\delta\phi=\int_0^1\left[I_a(z(\theta))\frac{de^a}{d\theta}
		+X_{ai}(z(\theta))\frac{d\psi^{ai}}{d\theta}\right]d\theta.
		\label{eq:exact_contact_increment}
	\end{equation}
	Expanding about the reference state $z_k$ gives the tangent relation
	$\delta\phi=I_{a,k}\delta e^a+X_{ai,k}\delta\psi^{ai}+\mathcal O(\|\Delta z\|^2)$.
	The residual $r_\phi$ used in the master action is this first-order tangent approximation. Thus the contact penalty enforces a \emph{linearized contact-increment condition}.
	
	In the standard continuum setting, local conservation reads $\delta e^a=-\nabla\cdot\delta\boldsymbol{\psi}^a$~\cite{GayBalmaz2017a,GayBalmaz2017b,Magnenet2020}. Integration by parts of $\delta\Phi_{\rm tot}=\int_\Omega I_a\delta e^a\,d\omega$ under insulated boundary conditions gives $\delta\Phi_{\rm tot}=\int_\Omega \nabla I_a\cdot\delta\boldsymbol{\psi}^a\,d\omega$. We define the representation-dependent kinetic driving-force increment by $\delta\mathbf{F}_a:=-\chi\nabla(\delta I_a)$. The contact coordinate $\mathbf X_a=\partial\phi/\partial\boldsymbol\psi^a$ and the gradient-induced force increment $\delta\mathbf F_a$ are distinct objects, and the present action does not impose $\mathbf X_{a,k}=\delta\mathbf F_a$.
	
	Along an irreversible trajectory generated by $\zeta$, the rate of change of the cumulated flow is $\dot{\boldsymbol{\psi}}^{a} = \mathbf{j}^a$ ($\delta\boldsymbol{\psi}^a = \mathbf{j}^a\delta t$)~\cite{Magnenet2020}.
	
	\subsection{Representation Convexity and Partial Legendre Transformation}
	In geometric thermodynamics, it is important to distinguish between two distinct operations~\cite{Salamon1984,Mrugala1991,Quevedo2007}: (i) an \emph{Energy--Entropy Representation Change}, which rearranges the coordinates of the equilibrium hypersurface and rescales the contact 1-form by $-1/T$, relating $g^W$ and $g^R$ conformally via $g^W = T \, g^R$; and (ii) a \emph{Total Legendre Transformation} on $\mathcal{T}_{\mathrm{ext}}$, which exchanges extensive and intensive coordinates~\cite{Arnold1980,Quevedo2007}:
	\begin{align}
		\phi &= \tilde{\phi} - \tilde{e}_a \tilde{I}^a - \sum_a \tilde{\boldsymbol{\psi}}_a\cdot\tilde{\mathbf{X}}^a, \nonumber \\
		e^a &= -\tilde{I}^a, \quad I_a = \tilde{e}_a, \quad \boldsymbol{\psi}^a = -\tilde{\mathbf X}^a, \quad \mathbf{X}_a = \tilde{\boldsymbol{\psi}}_a.
		\label{eq:total_legendre}
	\end{align}
	The canonical contact form is invariant under this full transformation; this is a statement about the contact geometry alone and does not imply that the master action is invariant under the full transformation. For the constitutive Hessian used below, we use the equilibrium-sector partial transformation, holding the transport coordinates fixed:
	\begin{align}
		\tilde\phi&=\phi-e^aI_a,\quad \tilde e_a=I_a,\quad \tilde I^a=-e^a,\nonumber\\
		\tilde\psi^{ai}&=\psi^{ai},\quad \tilde X_{ai}=X_{ai}.
		\label{eq:partial_legendre}
	\end{align}
	This partial transformation also preserves $\Theta_{\mathrm{ext}}$. Differentiating $\tilde I^a=-e^a$ with respect to $\tilde e_b=I_b$ gives the equilibrium-sector Hessian law for $T_{ab}=\partial I_a/\partial e^b$~\cite{Callen1985,Quevedo2007}:
	\begin{align}
		\tilde T^{ab}&:=\frac{\partial^2\tilde\phi}{\partial\tilde e_a\partial\tilde e_b}
		=\frac{\partial\tilde I^a}{\partial\tilde e_b}
		=-\frac{\partial e^a}{\partial I_b}=-(T^{-1})^{ab},\nonumber\\
		(\tilde T^{-1})_{ab}&=-T_{ab}.
		\label{eq:hessian_transform}
	\end{align}
	Let $\chi \in \{+1, -1\}$ denote the \emph{convexity sign} of the generating thermodynamic potential $\phi$~\cite{Callen1985,Rockafellar1970}:
	\begin{equation}
		\chi = \begin{cases}
			+1, & \text{convex potential (e.g., $\phi=u$)}, \\
			-1, & \text{concave potential (e.g., $\phi=s$)}.
		\end{cases}
		\label{eq:chi_def}
	\end{equation}
	Under this equilibrium-sector partial transformation, $\tilde{\chi}=-\chi$, so $g_{ab}\equiv\chi T_{ab}$ and $g^{ab}\equiv\chi(T^{-1})^{ab}$ are positive-definite wherever the chosen representation is strictly stable. For a general full Legendre transformation with nonzero mixed terms, the block relation in Eq.~\eqref{eq:hessian_transform} is not asserted. Within the partial equilibrium transformation, $\chi = \pm 1$ is spatially uniform.\newline The two quadratic forms used below are completions of the Tisza Hessian:
	\begin{align}
		\mathcal{Q}^{II}_{\mathrm{eq}}(\delta e, \delta I) &= \chi \, \delta e^a \delta I_a, \label{eq:QII_bilinear} \\
		\mathcal{Q}^I_{\mathrm{eq}}(\delta e, \delta I) &= \frac{\chi}{2}\left[ T_{ab}\delta e^a \delta e^b + (T^{-1})^{ab}\delta I_a \delta I_b \right]. \label{eq:QI_diagonal}
	\end{align}
	Under the equilibrium-sector partial Legendre transformation, the two forms are invariant in form, and their difference defines the positive semi-definite \emph{Legendre-dual, Fenchel--Young-type equilibrium constitutive defect}~\cite{Rockafellar1970,Mielke2011}:
	\begin{align}
		\mathcal D_{\rm eq}&:=\mathcal Q^I_{\rm eq}-\mathcal Q^{II}_{\rm eq}\nonumber\\
		&=\frac{\chi}{2}T_{ab}
		\left(\delta e^a-(T^{-1})^{ac}\delta I_c\right)\nonumber\\
		&\qquad\times
		\left(\delta e^b-(T^{-1})^{bd}\delta I_d\right)\ge0.
		\label{eq:GTD_defect}
	\end{align}
	Let $\Phi(x)=\frac12 x^{\mathsf T}(\chi T)x$, whose convex conjugate is $\Phi^*(p)=\frac12p^{\mathsf T}(\chi C)p$ with $C=T^{-1}$. Setting $x=\delta e$ and $p=\chi\delta I$ gives $\mathcal D_{\rm eq}=\Phi(x)+\Phi^*(p)-p^{\mathsf T}x$. Thus the quadratic defect is exactly a Fenchel--Young gap for the signed dual variable $p=\chi\delta I$.
	
	\begin{remark}[Metric form of the quadratic sectors]
		\label{rem:metric_form}
		Let $g_{ab}:=\chi T_{ab}$ and $g^{ab}:=\chi C^{ab}$. On the strictly stable region, both are positive definite and mutually inverse. Define $|\delta e|_g^2:=g_{ab}\delta e^a\delta e^b$ and $|\delta I|_{g^{-1}}^2:=g^{ab}\delta I_a\delta I_b$. Set $r_e:=\delta e-C\delta I$. Then
		\begin{align}
			\mathcal D_{\rm eq}
			&=\frac12|r_e|_g^2
			=\frac12|\delta e|_g^2
			+\frac12|\delta I|_{g^{-1}}^2
			-\chi\,\delta e^a\delta I_a.
			\label{eq:Deq_metric}
		\end{align}
		Under the equilibrium-sector partial Legendre transformation, $\widetilde{g}^{ab}=\widetilde{\chi}\,\widetilde{T}^{ab}=\chi(T^{-1})^{ab}=g^{ab}$, so the equilibrium squared norms are exchanged while the pairing $\chi\delta e^a\delta I_a$ is invariant. For $\chi=+1$, $g$ is the Weinhold metric in the energy representation; for $\chi=-1$, it is the Ruppeiner metric in the entropy representation. In the corresponding natural coordinates these satisfy $g^W=Tg^R$~\cite{Salamon1984,Mrugala1991}.
	\end{remark}
	
	\begin{proposition}[Full-defect Legendre self-duality]
		\label{prop:full_defect}
		Let $\phi(Z)$ be a $C^2$ potential on a domain where its symmetric full Hessian $\mathbb T_{AB}=\partial_A\partial_B\phi$ is nonsingular. Assume $\chi\mathbb T\succ0$. Define
		\[
		r^A:=\delta Z^A-\mathbb T^{-1,AB}\delta J_B,\qquad
		\mathcal D_{\mathrm{full}}:=\frac{\chi}{2}r^A\mathbb T_{AB}r^B.
		\]
		Under the locally defined full Legendre transformation $\widetilde Z_A=J_A$, $\widetilde J^A=-Z^A$, the defect is invariant, $\widetilde{\mathcal D}_{\mathrm{full}}=\mathcal D_{\mathrm{full}}$, and non-negative.
	\end{proposition}
	
	\begin{proof}
		The Legendre map gives $\widetilde{\mathbb T}=-\mathbb T^{-1}$, hence $\widetilde{\mathbb T}^{-1}=-\mathbb T$, and reverses the convexity sign, $\widetilde\chi=-\chi$. The transformed residual is
		$\widetilde r_A=\delta J_A-\mathbb T_{AB}\delta Z^B=-\mathbb T_{AB}r^B$. Therefore
		\[
		\widetilde{\mathcal D}_{\mathrm{full}}
		=\frac{-\chi}{2}\widetilde r_A(-\mathbb T^{-1})^{AB}\widetilde r_B
		=\frac{\chi}{2}r^A\mathbb T_{AB}r^B=\mathcal D_{\mathrm{full}}.
		\]
		Since $\chi\mathbb T\succ0$, there is a factorization $\chi\mathbb T=M^{\mathsf T}M$, and $\mathcal D_{\mathrm{full}}=\tfrac12\|Mr\|^2\geq0$.
	\end{proof}
	
	\begin{proposition}[Off-shell invariance of the equilibrium-storage-Lagrange sector]
		\label{prop:eq_storage_invariance}
		Consider $\mathcal A_{\rm eq} = \mathcal D_{\rm eq} + \frac{\chi}{2}\delta I^{\mathsf T} C\,\delta I + \lambda^{\mathsf T} r_e$. Under the partial Legendre transformation, with the affine law $\tilde\lambda = -C\lambda + \frac{\chi}{2}(\delta e + C\delta I)$, the sector is invariant off shell: $\widetilde{\mathcal A}_{\rm eq} = \mathcal A_{\rm eq}$.
	\end{proposition}
	
	\begin{proof}
		The residual transforms as $\tilde r_e=\delta I-(-T)(-\delta e)=-T r_e$. By symmetry of $T$ and $C=T^{-1}$,
		\[
		\widetilde{\mathcal D}_{\rm eq}
		=\frac{-\chi}{2}(-Tr_e)^{\mathsf T}(-C)(-Tr_e)
		=\frac{\chi}{2}r_e^{\mathsf T}T r_e
		=\mathcal D_{\rm eq}.
		\]
		The transformed storage term is $\widetilde S_{\rm st}=\frac{\chi}{2}\delta e^{\mathsf T}T\delta e$. Hence
		\begin{equation}
			\widetilde S_{\rm st}-S_{\rm st}
			=\frac{\chi}{2}(\delta e+C\delta I)^{\mathsf T}T r_e.
			\label{eq:storage_diff_eq}
		\end{equation}
		The multiplier law gives $T\tilde\lambda=-\lambda+\frac{\chi}{2}T(\delta e+C\delta I)$, so that
		\[
		\tilde\lambda^{\mathsf T}\tilde r_e-\lambda^{\mathsf T}r_e
		=-\frac{\chi}{2}(\delta e+C\delta I)^{\mathsf T}T r_e,
		\]
		which exactly cancels Eq.~\eqref{eq:storage_diff_eq}. Adding the three contributions proves $\widetilde{\mathcal A}_{\rm eq}=\mathcal A_{\rm eq}$ off shell.
	\end{proof}
	
	\begin{remark}[Scope of full Legendre invariance]
		\label{rem:legendre_scope}
		The full Legendre transformation does not prescribe a law for the flux-rate increments, multipliers, or reference scale. Off-shell invariance of $\mathcal S_{\rm aug}$ under the full transformation is not claimed here and is discussed in Sec.~IV.
	\end{remark}
	
	\section{A Constrained Mixed Variational Principle}
	\label{sec:master_action}
	
	\subsection{Constitutive Residuals and the Master Action}
	In the remainder of this section we identify each flux-family index with the corresponding conserved-variable index. The coefficients $T_{ab}$, $C^{ab}=(T^{-1})^{ab}$, and $L^{ab}$ are evaluated at the known reference state $t_k$ and held fixed during variations over one increment. We assume $\chi T\succ0$ and $L=L^{\mathsf T}\succ0$ on the admissible stable region.
	
	Let $C^{ab}:=(T^{-1})^{ab}$. We retain the non-negative defects
	\begin{align}
		\mathcal D_{\rm eq}&:=\frac{\chi}{2}T_{ab}
		\bigl(\delta e^a-C^{ac}\delta I_c\bigr)
		\bigl(\delta e^b-C^{bd}\delta I_d\bigr),\label{eq:Deq_master}\\
		\mathcal D_{\rm diss}&:=\frac14(L^{-1})_{ab}
		\bigl(\delta\mathbf j^a-L^{ac}\delta\mathbf F_c\bigr)\cdot
		\bigl(\delta\mathbf j^b-L^{bd}\delta\mathbf F_d\bigr),
		\label{eq:Ddiss_master}
	\end{align}
	where $\delta\mathbf F_a=-\chi\nabla(\delta I_a)$.
	
	Let $\varsigma_T:=\lambda_{\min}(\chi T)>0$ on the stable region, so that $\chi T\succeq\varsigma_T I$. Then the equilibrium defect controls the constitutive residual by
	\begin{equation}
		\|r_e\|^2\le\frac{2\mathcal D_{\rm eq}}{\varsigma_T}.
		\label{eq:Deq_residual_bound}
	\end{equation}
	The bound is informative only when $\delta e$ is obtained independently of $C\delta I$ (for example from a nonlinear constitutive update), in which case $\mathcal D_{\rm eq}$ measures the deviation from the frozen linearized Tisza law; on stationary points of the master action $r_e=0$ and the bound is trivial.
	
	The dissipative defect is a Fenchel--Young gap. Define
	\[
	\Psi(j):=\frac14(L^{-1})_{ab}j^a\cdot j^b,\qquad
	\Psi^*(p):=L^{ab}p_a\cdot p_b.
	\]
	Setting $p_a=\delta\mathbf F_a/2$ gives
	\begin{equation}
		\mathcal D_{\rm diss}=\Psi(\delta\mathbf j)+\Psi^*(\delta\mathbf F/2)-\frac12\delta\mathbf j^a\cdot\delta\mathbf F_a\geq0.
		\label{eq:Ddiss_FY}
	\end{equation}
	Thus $\mathcal D_{\rm diss}$ vanishes exactly when $\delta\mathbf j^a=L^{ab}\delta\mathbf F_b$. The representation of the dissipative gap as a Fenchel--Young inequality is the incremental counterpart of the Onsager--Machlup least-dissipation functional~\cite{OnsagerMachlup1953} and of the modern metric-space formulation of gradient flows and large deviations~\cite{MielkePeletierRenger2014}.
	
	\begin{proposition}[Nonlinear dissipative potentials]
		\label{prop:nonlinear_dissipation}
		Let $\Psi:\mathbb{R}^{nd}\to\mathbb{R}\cup\{+\infty\}$ be proper, convex and lower-semicontinuous, with convex conjugate $\Psi^*(p)=\sup_j\{p\cdot j-\Psi(j)\}$. Define
		\[
		\mathcal D_{\rm diss}^{\rm NL}(\delta j,\delta F):=\Psi(\delta j)+\Psi^*(\delta F/2)-\tfrac12\delta j\cdot\delta F.
		\]
		Then $\mathcal D_{\rm diss}^{\rm NL}\ge0$ for all $(\delta j,\delta F)$, with equality if and only if $\delta j\in\partial\Psi^*(\delta F/2)$. In the quadratic case $\Psi(\delta j)=\frac14(L^{-1})_{ab}\delta j^a\cdot\delta j^b$, this reduces to Eq.~\eqref{eq:Ddiss_FY} with $\nabla\Psi^*(\delta F/2)=L\delta F$, recovering the Onsager law.
	\end{proposition}
	
	\begin{proof}
		Non-negativity is the Fenchel--Young inequality. Equality holds if and only if $\delta F/2\in\partial\Psi(\delta j)$, equivalent to $\delta j\in\partial\Psi^*(\delta F/2)$ under standard conjugacy.
	\end{proof}
	
	Define the residuals
	\begin{align}
		r_\phi&:=\delta\phi-I_{a,k}\delta e^a
		-\mathbf X_{a,k}\cdot(\mathbf j_k^a+\delta\mathbf j^a)\delta t,\nonumber\\
		r_e^a&:=\delta e^a-C^{ab}\delta I_b,\qquad
		\mathbf r_j^a:=\delta\mathbf j^a-L^{ab}\delta\mathbf F_b.
		\label{eq:master_residuals}
	\end{align}
	Here $I_{a,k}$, $\mathbf X_{a,k}$ and $\mathbf j_k^a$ are reference-state quantities at $t_k$, and $\delta\boldsymbol\psi^a=(\mathbf j_k^a+\delta\mathbf j^a)\delta t$. The proposed \emph{constrained mixed master action} is
	\begin{align}
		\mathcal S_{\rm aug}&[\delta\phi,\delta e,\delta\mathbf j,\delta I,\lambda,\boldsymbol\mu]
		=\int_\Omega\Bigg\{\frac{r_\phi^2}{2\phi_0}
		+\mathcal D_{\rm eq}+\mathcal D_{\rm diss}\,\delta t\nonumber\\
		&\quad+\frac12\chi C^{ab}\delta I_a\delta I_b
		+\left(\mathbf j_k^a+\frac12 L^{ab}\delta\mathbf F_b\right)
		\cdot\delta\mathbf F_a\,\delta t\nonumber\\
		&\quad+\lambda_a r_e^a+\boldsymbol\mu_a\cdot\mathbf r_j^a
		\Bigg\}\,d\omega.
		\label{eq:master_action_single}
	\end{align}
	Here $\phi_0>0$ is a fixed reference potential-density scale with the same units as $r_\phi$. The multiplier fields are assigned units such that each product $\lambda_a r_e^a$ and $\boldsymbol\mu_a\cdot\mathbf r_j^a$ has the same units as the potential-density terms in the integrand. Because each quadratic defect and its first variation vanish on its zero-residual set, the defects provide a non-negative off-shell Fenchel--Young gap without altering the first-order on-shell stationarity equations enforced by the multipliers.
	
	\begin{remark}[Role of the contact sector]
		\label{rem:contact_passive}
		Since $\delta\phi$ appears only in $r_\phi$, the contact stationarity condition algebraically determines $\delta\phi$ along the tangent Legendrian space without feeding back into the transport equations. The contact sector is thus decoupled from the balance law, which is generated entirely by the storage--transport sector together with the constitutive constraints.
	\end{remark}
	
	\subsection{Stationarity and Exact Elimination}
	We impose $\delta I_a=0$ on Dirichlet portions of the boundary and the natural insulated condition $\mathbf n\cdot(\mathbf j_k^a+\delta\mathbf j^a)=0$ on portions where the intensive increment is free.
	
	Variation with respect to the independent potential increment and the multipliers gives
	\begin{align}
		\frac{\delta\mathcal S_{\rm aug}}{\delta(\delta\phi)}&=\frac{r_\phi}{\phi_0}=0,\label{eq:single_EL_phi}\\
		\frac{\delta\mathcal S_{\rm aug}}{\delta\lambda_a}&=r_e^a=0,\label{eq:single_EL_e}\\
		\frac{\delta\mathcal S_{\rm aug}}{\delta\boldsymbol\mu_a}&=\mathbf r_j^a=0.\label{eq:single_EL_j}
	\end{align}
	Thus the linearized contact-increment constraint, the linearized Tisza relation $\delta e^a=C^{ab}\delta I_b$, and the Onsager relation $\delta\mathbf j^a=L^{ab}\delta\mathbf F_b$ hold at a stationary point.
	
	On the common constraint set $r_\phi=r_e=\mathbf r_j=0$, the first variations of the contact penalty and of both quadratic defects vanish. The variations with respect to the extensive and flux increments therefore give
	\begin{align}
		\left.\frac{\delta\mathcal S_{\rm aug}}{\delta(\delta e^a)}\right|_{\rm constraints}&=\lambda_a=0,\label{eq:lambda_stationarity}\\
		\left.\frac{\delta\mathcal S_{\rm aug}}{\delta(\delta\mathbf j^a)}\right|_{\rm constraints}&=\boldsymbol\mu_a=0.\label{eq:mu_stationarity}
	\end{align}
	The remaining variation with respect to $\delta I_a$ reduces to
	\begin{align}
		0&=\chi C^{ab}\delta I_b+\chi\delta t\,\nabla\!\cdot\mathbf j_k^a
		-\delta t\,\nabla\!\cdot\bigl(L^{ab}\nabla\delta I_b\bigr).
		\label{eq:single_EL_I}
	\end{align}
	Using $\delta\mathbf j^a=L^{ab}\delta\mathbf F_b=-\chi L^{ab}\nabla\delta I_b$ and $\delta e^a=C^{ab}\delta I_b$, Eq.~\eqref{eq:single_EL_I} is equivalent to
	\begin{equation}
		\boxed{\delta e^a+\delta t\,\nabla\!\cdot(\mathbf j_k^a+\delta\mathbf j^a)=0.}
		\label{eq:implicit_transport}
	\end{equation}
	Therefore, under the frozen-coefficient and boundary assumptions stated above, the linearized contact condition, Tisza relation, Onsager relation, and implicit balance equation arise as simultaneous stationarity conditions of one augmented scalar action.
	
	\begin{proposition}[Exact elimination of the auxiliary fields]
		\label{prop:exact_elimination}
		Let the coefficients be frozen as above, and let the discrete trial spaces be compatible in the sense that the reconstructions below belong to the corresponding spaces. A stationary point of the augmented action is equivalent to a solution $\delta I_h\in V_h$ of
		\begin{equation}
			\begin{aligned}
				0={}&\int_\Omega\Big[
				\chi C^{ab}\delta I_{b,h}v_{a,h}
				-\chi\delta t\,\mathbf j_k^a\cdot\nabla v_{a,h}\\
				&\qquad+\delta t\,L^{ab}\nabla\delta I_{b,h}\cdot\nabla v_{a,h}
				\Big]d\omega,\quad \forall\,v_h\in V_h,
			\end{aligned}
			\label{eq:reduced_weak_galerkin}
		\end{equation}
		together with the algebraic reconstruction
		\begin{align}
			\delta e_h^a&=C^{ab}\delta I_{b,h},&
			\delta\mathbf j_h^a&=-\chi L^{ab}\nabla\delta I_{b,h},\nonumber\\
			\delta\phi_h&=I_{a,k}\delta e_h^a+\nonumber\\
			&\quad\mathbf X_{a,k}\cdot(\mathbf j_k^a+\delta\mathbf j_h^a)\delta t,\nonumber\\
			\lambda_{a,h}&=0,&\boldsymbol\mu_{a,h}&=0.
			\label{eq:exact_reconstruction}
		\end{align}
		Thus, the multiplier fields need not be solved independently.
	\end{proposition}
	
	\begin{proof}
		The variations with respect to $\lambda_h$ and $\boldsymbol\mu_h$ impose $r_{e,h}=0$ and $\mathbf r_{j,h}=0$, giving the first line of Eq.~\eqref{eq:exact_reconstruction}. Variation with respect to $\delta\phi_h$ imposes $r_{\phi,h}=0$, giving the contact reconstruction. On this common constraint set, the first variations of the quadratic residual penalties vanish, and the variations with respect to $\delta e_h$ and $\delta\mathbf j_h$ give $\lambda_h=0$ and $\boldsymbol\mu_h=0$. The remaining $\delta I_h$ variation is precisely the reduced weak balance equation. Conversely, a solution of that reduced equation, lifted by Eq.~\eqref{eq:exact_reconstruction}, satisfies all six stationarity blocks.
	\end{proof}
	
	\begin{remark}[Interpretation and numerical scope]
		\label{rem:action_reduction}
		Proposition~\ref{prop:exact_elimination} makes the role of the master action precise: the augmented functional packages the contact relation, constitutive constraints, Legendre-dual defects, and storage--transport balance into one scalar variational statement, and the multipliers enforce the same residual constraints whose quadratic defects vanish on shell, so the defects provide non-negative off-shell defect measures without changing the stationary balance equation. The proposition provides an algebraic elimination and reconstruction procedure rather than a modification of the underlying evolution equation. The nonlinear $DG_0$--$RT_0$ benchmark below is designed specifically to test the conservative nonlinear balance--constitutive subsystem with state-dependent conductivity, whereas a full saddle-point discretization retaining all six multiplier and contact blocks of $\mathcal S_{\rm aug}$ as independent unknowns is left for future work.
	\end{remark}
	
	\begin{proposition}[Intrinsic discrete conservation in the mixed formulation]
		\label{prop:mixed_conservation}
		Let $V_h$ be the scalar test space used for the discrete balance equation and assume that the constant function $1$ belongs to $V_h$. Any solution of the conservative discrete balance equation with zero total boundary flux then satisfies $\int_\Omega\delta e^a\,d\omega=0$ for each conserved component $a$.
	\end{proposition}
	
	\begin{proof}
		Test the weak balance equation for each component $a$ with the constant function $\varphi\equiv 1$. The diffusive Onsager term vanishes because $\nabla\varphi=0$; the remaining identity is $\int_\Omega\delta e^a\,d\omega=-\delta t\int_{\partial\Omega}(\mathbf j_k^a+\delta\mathbf j^a)\cdot\mathbf n\,dS=0$.
	\end{proof}
	
	\section{Nonlinear Extension via Fenchel--Young and Bregman Divergences}
	\label{sec:bregman}
	
	Let $x=\delta e$ and $\xi=\chi\,\delta I$, and introduce the centered function
	\begin{equation}
		W(x)=\chi\big[\phi(e_k+x)-\phi(e_k)-I(e_k)\cdot x\big].
		\label{eq:W_def}
	\end{equation}
	Assume that $W$ is proper, lower-semicontinuous, strictly convex, with positive-definite Hessian where evaluated. Its Legendre--Fenchel conjugate is
	\begin{equation}
		W^*(\xi)=\sup_x\{\xi\cdot x-W(x)\}.
		\label{eq:W_conjugate}
	\end{equation}
	The associated nonlinear Fenchel--Young defect is
	\begin{equation}
		\begin{aligned}
			\mathcal D_{\rm FY}^{\rm NL}(x,\xi)
			&=W(x)+W^*(\xi)-\xi\cdot x\\
			&=D_W\bigl(x,x^*(\xi)\bigr)\geq0.
		\end{aligned}
		\label{eq:Deq_Bregman}
	\end{equation}
	where $x^*(\xi)=\nabla W^*(\xi)$ and $D_W$ is the Bregman divergence~\cite{Bregman1967}. Thus the Fenchel--Young gap is a Bregman divergence in the primal space; it is not a divergence with the dual variable $\xi$ used directly as a primal point.
	
	\paragraph{Self-duality.} Under the equilibrium-sector partial Legendre transformation, use $\tilde\chi=-\chi$ and define
	\begin{equation}
		\begin{aligned}
			\tilde x&=\chi \xi, & \tilde \xi&=\chi x,\\
			\tilde W(y)&=W^*(\chi y), & \tilde W^*(z)&=W(\chi z).
		\end{aligned}
		\label{eq:nonlinear_dual_map}
	\end{equation}
	These definitions are consistent because $\chi\in\{+1,-1\}$ and the conjugate of $y\mapsto W^*(\chi y)$ is $z\mapsto W(\chi z)$. Direct substitution gives
	\begin{align}
		\tilde{\mathcal D}_{\rm FY}^{\rm NL}
		&=\tilde W(\tilde x)+\tilde W^*(\tilde \xi)
		-\tilde \xi\cdot\tilde x\nonumber\\
		&=W^*(\xi)+W(x)-\xi\cdot x
		=\mathcal D_{\rm FY}^{\rm NL}.
		\label{eq:nonlinear_defect_selfdual}
	\end{align}
	This is an off-shell identity for the defect itself.
	
	A nonlinear constitutive residual can be defined by $r_{\rm NL}:=x-\nabla W^*(\xi)$. With the mean Hessian $\bar H$ along the segment from $x^*$ to $x$, the transformed residual satisfies $\tilde r_{\rm NL}=-\chi\bar H r_{\rm NL}$. Whenever $\bar H$ is invertible, the residual pairing can be preserved by the dual multiplier law $\tilde\lambda=-\chi\bar H^{-1}\lambda$ for symmetric $\bar H$. This law transforms only the nonlinear constitutive-constraint term. The additional affine term in Proposition~\ref{prop:eq_storage_invariance} compensates the separate quadratic storage term; it is not generated by the constraint pairing alone.
	
	\paragraph{Scope of the nonlinear extension.}
	The master action in Eq.~\eqref{eq:master_action_single} remains the quadratic, frozen-coefficient action used in the numerical benchmarks. The construction above defines a nonlinear defect and its duality properties; it does not silently replace the quadratic residual and storage terms. For the natural nonlinear storage--defect block with $S_{\rm st}=W^*(\xi)$, the mismatch on the constraint $\xi=\nabla W(x)$ is
	\begin{equation}
		W(x)-W^*(\xi)=2W(x)-\xi\cdot x.
		\label{eq:nonlinear_storage_mismatch}
	\end{equation}
	This vanishes for centered functions that are positively homogeneous of degree two (by Euler's identity) but not for generic nonlinear $W$. This establishes the non-invariance of this natural block for generic non-quadratic potentials; a fully nonlinear augmented action that preserves the physical balance law and full Legendre invariance is discussed among the open problems at the end.
	
	\section{Conformal Rescaling and Discrete Energy Conservation in Heat Conduction}
	\label{sec:heat}
	
	\subsection{Canonical Entropy Picture and the Energy Representation}
	We apply the formulation to Fourier heat conduction in the canonical entropy representation~\cite{Callen1985,DeGroot1984,Magnenet2020}, with $\phi = s(u)$ ($\chi = -1$), $e_1 = u$, and $I_1 = 1/T$. For this rigid-conductor benchmark, \(c_{\mathrm{eff}} \equiv du/dT\) is a prescribed constant.
	
	By concavity of entropy, $T_{uu}=-1/(c_{\mathrm{eff}}T_k^2)<0$ and $(T_{uu})^{-1}=-c_{\mathrm{eff}}T_k^2<0$. The first-order relation $\delta I_1 = -\delta T_k/T_k^2 + \mathcal{O}(\delta T_k^2)$ inserted into $r_e=0$ gives $\delta u = +c_{\mathrm{eff}} \delta T_k$ at first order. The Onsager mobility is \(L_{uu} = \kappa T_k^2 > 0\)~\cite{DeGroot1984}, giving $\delta\mathbf q_k = \kappa T_k^2\nabla(\delta I_1)$.
	
	In the temperature-variable benchmark, $\kappa$ is constant and the Fourier-flux increment is $\delta\mathbf q_k=-\kappa\nabla(\delta T_k)$. This must be distinguished from direct substitution into the force-based Onsager increment: using $\delta I_1\simeq-\delta T_k/T_k^2$ in $\delta\mathbf q_{\rm Ons}=\kappa T_k^2\nabla(\delta I_1)$ gives
	\[
	\delta\mathbf q_{\rm Ons}=-\kappa\nabla(\delta T_k)+2\kappa\frac{\delta T_k}{T_k}\nabla T_k+\mathcal O(\delta T_k^2),
	\]
	because the spatial derivative also acts on $T_k^{-2}$. Thus the temperature-variable functional analyzed next is not obtained by pointwise substitution into the entropy-representation functional; it is the energy-representation storage--transport sector of the same incremental construction, supplemented by the explicit entropy-production source in Eq.~\eqref{eq:S_energy_picture}. Proposition~\ref{prop:energy_picture} establishes its relation to Eq.~(54) of Ref.~\cite{Magnenet2020}. If $\kappa=\kappa(T)$ is varied fully, the Fourier increment also contains $-\kappa'(T_k)\delta T_k\nabla T_k$, and the drift identity must be rederived. The nonlinear $\kappa(T)=1+T^2$ calculation reported later is a separate conservative balance/constitutive benchmark.
	
	\subsection{Weighted Drift Identity}
	In the energy representation, the three-term incremental functional of Ref.~\cite{Magnenet2020}, with the local reference temperature $T_k(x)$ retained, is:
	\begin{equation}
		\begin{aligned}
			\mathcal T_{\rm MS}[\delta T_k]=\int_\Omega\Bigg[
			&c_{\rm eff}\frac{(\delta T_k)^2}{2T_k}\\
			&-\frac{1}{T_k}\left(\mathbf q_k+\frac12\delta\mathbf q_k\right)
			\cdot\nabla(\delta T_k)\,\delta t\\
			&-\frac{|\mathbf q_k|^2}{\kappa T_k^2}
			\delta T_k\,\delta t\Bigg]d\omega.
		\end{aligned}
		\label{eq:MS_eq54}
	\end{equation}
	Let $\mathbf q_{k+1}=\mathbf q_k+\delta\mathbf q_k$ and impose $\mathbf n\cdot\mathbf q_{k+1}=0$. For constant or frozen $\kappa$, variation of Eq.~\eqref{eq:MS_eq54} gives
	\begin{align}
		\frac{\delta\mathcal T_{\rm MS}}{\delta(\delta T_k)}
		&=\frac{1}{T_k}\Bigg[c_{\rm eff}\delta T_k
		+\delta t\nabla\cdot\mathbf q_{k+1}\nonumber\\
		&\qquad+\delta t\frac{\kappa\nabla T_k\cdot\nabla(\delta T_k)}{T_k}
		\Bigg]=0.
		\label{eq:EL_MS54_3term}
	\end{align}
	Integration by parts of $-\mathbf q_k\cdot\nabla(\delta T_k)/T_k$ produces the background term $+\kappa|\nabla T_k|^2/T_k^2$; the third term in Eq.~\eqref{eq:MS_eq54} cancels it. For this heat-conduction benchmark, define the total thermal energy at time level $k$ by
	\begin{equation}
		\begin{aligned}
			E_k&:=\int_\Omega c_{\rm eff}T_k\,d\omega,\\
			\delta E_k&:=E_{k+1}-E_k
			=\int_\Omega c_{\rm eff}\delta T_k\,d\omega.
		\end{aligned}
		\label{eq:thermal_energy_definition}
	\end{equation}
	The remaining cross-gradient term yields
	\begin{equation}
		\begin{aligned}
			\delta E_k^{\mathrm{(3\text{-}term)}}
			&:=\int_\Omega c_{\rm eff}\delta T_k\,d\omega\\
			&=-\delta t\int_\Omega
			\frac{\kappa\nabla T_k\cdot\nabla(\delta T_k)}{T_k}\,d\omega.
		\end{aligned}
		\label{eq:energy_gain_3term}
	\end{equation}
	In our benchmark, $\nabla T_k\cdot\nabla(\delta T_k)<0$, so the drift is positive. For general data, the sign is not guaranteed.
	
	For $0<T_{\mathrm{min}}\leq T_k\leq T_{\mathrm{max}}<\infty$, the quadratic functional is coercive and strictly convex on $H^1(\Omega)$: its second variation is
	\[
	\int_\Omega\frac{c_{\rm eff}}{T_k}v^2\,d\omega
	+\delta t\,\kappa\int_\Omega\frac{|\nabla v|^2}{T_k}\,d\omega>0
	\quad (v\neq0).
	\]
	Hence it admits a unique minimizer $\delta T_k$, and the insulated condition is its natural boundary condition.
	
	The weighted structure behind Eq.~\eqref{eq:energy_gain_3term} is made explicit by multiplying the bracket in Eq.~\eqref{eq:EL_MS54_3term} by \(\delta T_k/T_k\) (equivalently, the full equation by \(\delta T_k\)). Integrating by parts with the Neumann conditions gives
	\begin{equation}
		\begin{aligned}
			\int_\Omega\frac{c_{\rm eff}\delta T_k^2}{T_k}\,d\omega
			&+\delta t\,\kappa\int_\Omega\frac{|\nabla\delta T_k|^2}{T_k}\,d\omega\\
			&=-\delta t\int_\Omega\frac{R_k}{T_k}\,\delta T_k\,d\omega.
		\end{aligned}
		\label{eq:weighted_identity}
	\end{equation}
	where \(R_k:=-\kappa\Delta T_k\). The two terms on the left are non-negative and the right-hand side involves \(R_k/T_k\). The cross-gradient term in the unweighted equation is precisely what makes the linearized operator self-adjoint in the weighted inner product \(L^2(\Omega;\,d\omega/T_k)\).
	
	\begin{proposition}[Weighted drift identity and per-step bound]
		\label{prop:drift_bound}
		Assume $d\le 3$, \(T_{\min}\le T_k\le T_{\max}\) with \(T_{\min}>0\), \(\kappa>0\) constant, \(T_k\in H^2(\Omega)\) with \(\partial_\nu T_k=0\) on \(\partial\Omega\), and $\vartheta_k:=\nabla\cdot(\kappa\nabla\ln T_k)\in L^2(\Omega)$. Write \(R_k:=-\kappa\Delta T_k\) and \(m_0:=c_{\rm eff}/T_{\max}\). Then
		\begin{equation}
			\|\delta T_k\|_{L^2}
			\le\frac{\delta t}{m_0}\left\|\frac{R_k}{T_k}\right\|_{L^2},
			\label{eq:sharp_bound_dT}
		\end{equation}
		and the per-step drift satisfies
		\begin{equation}
			\begin{aligned}
				\bigl|\delta E_k^{\mathrm{(3\text{-}term)}}\bigr|
				&\leq\frac{\delta t^2}{m_0}\,\|\vartheta_k\|_{L^2}\\
				&\qquad\times\left\|\frac{R_k}{T_k}\right\|_{L^2}.
			\end{aligned}
			\label{eq:sharp_drift}
		\end{equation}
		The per-step drift is therefore \(O(\delta t^2)\) under \(H^2\)-regularity of \(T_k\) and $\vartheta_k\in L^2$, with the explicit constant $m_0^{-1}\|\vartheta_k\|_{L^2}\|R_k/T_k\|_{L^2}$. The identity \eqref{eq:weighted_identity} holds for every \(\delta t>0\).
	\end{proposition}
	
	\begin{proof}
		Dropping the non-negative gradient term in Eq.~\eqref{eq:weighted_identity} and using \(c_{\rm eff}/T_k\ge m_0\), Cauchy--Schwarz gives Eq.~\eqref{eq:sharp_bound_dT}. The drift identity \eqref{eq:energy_gain_3term} may be rewritten as \(\delta E_k^{\mathrm{(3\text{-}term)}}=\delta t\int_\Omega \vartheta_k\,\delta T_k\,d\omega\) after one further integration by parts, since \(\nabla\cdot(\kappa\nabla T_k/T_k)=\vartheta_k\) and \(\partial_\nu T_k=0\); Cauchy--Schwarz and Eq.~\eqref{eq:sharp_bound_dT} then give Eq.~\eqref{eq:sharp_drift}.
	\end{proof}
	
	\begin{proposition}[Conditional cumulative drift over a fixed time interval]
		\label{prop:cumulative_drift}
		Let $\Omega\subset\mathbb R^d$ be bounded and $C^{1,1}$ with $d\leq3$, and suppose the assumptions of Proposition~\ref{prop:drift_bound} hold at every step $k=0,\ldots,N-1$, with $N\delta t=T_f$. Here the $H^2$ norm is taken to be the equivalent Neumann graph norm $\|T\|_{H^2}:=\|T\|_{L^2}+\|\nabla T\|_{L^2}+\|\Delta T\|_{L^2}$ for fields satisfying $\partial_\nu T=0$. If, in addition, $\sup_{0\leq k<N}\|T_k\|_{H^2(\Omega)}\leq M_2<\infty$, then with $C_{\Omega}$ the embedding constant for $H^2(\Omega)\hookrightarrow W^{1,4}(\Omega)$ under this norm and
		\[
		M_*:=\frac{\kappa^2M_2}{T_{\min}}
		\left(\frac{M_2}{T_{\min}}+
		\frac{C_{\Omega}^2M_2^2}{T_{\min}^2}\right),
		\]
		the cumulative energy drift satisfies
		\begin{equation}
			|E_N-E_0|\leq \frac{T_f M_*}{m_0}\,\delta t.
			\label{eq:cumulative_drift}
		\end{equation}
		Thus the cumulative drift is $O(\delta t)$ as $\delta t\to0$ for fixed $T_f$, conditional on the uniform $H^2$ and positive-temperature bounds.
	\end{proposition}
	
	\begin{proof}
		Use the Neumann graph norm
		\[
		\|T\|_{H^2}:=\|T\|_{L^2}+\|\nabla T\|_{L^2}+\|\Delta T\|_{L^2}.
		\]
		Then
		\[
		\left\|\frac{R_k}{T_k}\right\|_{L^2}
		\leq\frac{\kappa M_2}{T_{\min}}.
		\]
		Since
		\[
		\vartheta_k=\kappa\left(\frac{\Delta T_k}{T_k}
		-\frac{|\nabla T_k|^2}{T_k^2}\right)
		\]
		and
		\[
		\|\nabla T_k\|_{L^4}\leq C_{\Omega}\|T_k\|_{H^2},
		\]
		we obtain
		\[
		\|\vartheta_k\|_{L^2}
		\leq\kappa\left(\frac{M_2}{T_{\min}}
		+\frac{C_{\Omega}^2M_2^2}{T_{\min}^2}\right).
		\]
		Combining these estimates with Eq.~\eqref{eq:sharp_drift} gives
		\[
		|\delta E_k^{\mathrm{(3\text{-}term)}}|
		\leq\frac{\delta t^2M_*}{m_0}
		\]
		at every step. Summing the increments and using the triangle inequality,
		\[
		|E_N-E_0|\leq\sum_{k=0}^{N-1}
		|\delta E_k^{\mathrm{(3\text{-}term)}}|
		\leq N\frac{\delta t^2M_*}{m_0}
		=\frac{T_fM_*}{m_0}\,\delta t,
		\]
		which is Eq.~\eqref{eq:cumulative_drift}.
	\end{proof}
	
	\begin{proposition}[One-dimensional a priori stability and cumulative drift]
		\label{prop:1d_weighted_stability}
		Let $\Omega=(0,L)$, $c:=c_{\rm eff}>0$ and $\kappa>0$ constant, and consider the insulated weighted update
		\begin{equation}
			\begin{aligned}
				\frac{c}{\delta t}v_k-\kappa T_{k+1}''
				&=-\kappa b_kv_k',\\
				v_k&:=T_{k+1}-T_k,\qquad b_k:=(\ln T_k)'.
			\end{aligned}
			\label{eq:1d_weighted_step}
		\end{equation}
		with $T_k'(0)=T_k'(L)=0$. Assume $T_0\in H^2(0,L)$, $T_0'(0)=T_0'(L)=0$, $T_0\geq T_->0$, and set $G_0:=\|T_0'\|_{L^2}$, $H_0:=\|T_0''\|_{L^2}$. There exist constants $C_L$ (depending only on $L$) and a sufficient time-step threshold $\delta t_0>0$ such that, for $0<\delta t\leq\delta t_0$ and every fixed $T_f=N\delta t$, the update has a unique $H^2$ solution at each step, and there are constants $M_T,C_2,C_P$ independent of $\delta t$ and $N$ with
		\begin{align}
			&T_k(x)\geq T_-,\quad
			\sup_{0\leq k\leq N}\|T_k'\|_{L^2}^2\leq G_0^2,
			\label{eq:1d_H1_uniform}\\
			&\frac14\sum_{k=0}^{N-1}\|v_k'\|_{L^2}^2
			+\frac{\kappa\delta t}{c}
			\sum_{k=0}^{N-1}\|T_{k+1}''\|_{L^2}^2
			\leq G_0^2,
			\label{eq:1d_L2H2}\\
			&\sup_{0\leq k\leq N}\|T_k\|_{L^\infty}\leq M_T.
			\label{eq:1d_upper_temperature}
		\end{align}
		The cumulative energy drift obeys
		\begin{equation}
			\max_{0\leq n\leq N}|E_n-E_0|
			\leq \frac{C_P}{m_0}\,\delta t,
			\qquad m_0:=\frac{c}{M_T}.
			\label{eq:1d_cumulative_drift}
		\end{equation}
	\end{proposition}
	
	The proof is given in Appendix~\ref{app:stability_1d}.
	
	\begin{remark}[Dimension and discretization scope]
		\label{rem:uniform_regularity}
		Proposition~\ref{prop:1d_weighted_stability} gives the a priori estimates only in one spatial dimension. Proposition~\ref{prop:cumulative_drift} gives a conditional cumulative-drift statement in dimensions two and three, provided a uniform-in-step $H^2$ bound and positive-temperature bounds are available. Mesh-uniform stability for a fully discrete Galerkin realization remains a separate open question.
		
		The one-dimensional proof controls $b_k=(\ln T_k)'$ in $L^\infty$ through the estimate in Eq.~\eqref{eq:1d_B_bound}, which applies $H^1(0,L)\hookrightarrow L^\infty(0,L)$ to $T_k'$. In dimensions two and three, the corresponding $H^1$ control of $\nabla T_k$ does not give an $L^\infty$ bound. Consequently, the absorption condition used in the proof, and hence the derived uniform $H^1$ and discrete $L^2(0,T_f;H^2)$ estimates, do not follow in those dimensions from the present argument. Establishing suitable higher-dimensional coefficient bounds or a different estimate is left open.
	\end{remark}
	
	\begin{proposition}[Energy-picture reduction and conformal-factor drift]
		\label{prop:energy_picture}
		Consider the energy representation $\chi=+1$ with $\phi=u(s)$, $e^1=s$, $I_1=T$, and $ds=c_{\rm eff}\,dT/T$. At the reference state, $T_{ss}=T_k/c_{\rm eff}$ and $C^{ss}=c_{\rm eff}/T_k$. Let $\kappa>0$ be constant, set
		\[
		\mathbf q_k=-\kappa\nabla T_k,\quad
		\mathbf j_{s,k}:=\frac{\mathbf q_k}{T_k},\quad
		L:=\frac{\kappa}{T_k},
		\]
		and use the frozen-temperature entropy-flux increment and force increment
		\[
		\delta\mathbf j_s:=\frac{\delta\mathbf q}{T_k},\quad
		\delta\mathbf F:=-\nabla\delta T,\quad
		\delta\mathbf q:=-\kappa\nabla\delta T.
		\]
		Define $\sigma_k:=|\mathbf q_k|^2/(\kappa T_k^2)$. For insulated boundary conditions, consider the energy-representation storage--transport sector of the augmented action, supplemented by this prescribed entropy-production source:
		\begin{equation}
			\begin{aligned}
				\mathcal S^{(s)}_{\rm red}[\delta T]
				=\int_\Omega\Bigg[
				&\frac12 g^{ss}(\delta T)^2
				+\mathbf j_{s,k}\!\cdot\!\delta\mathbf F\,\delta t\\
				&+\frac12L|\delta\mathbf F|^2\delta t
				-\sigma_k\delta T\,\delta t
				\Bigg]d\omega.
			\end{aligned}
			\label{eq:S_energy_picture}
		\end{equation}
		Then:
		\begin{enumerate}
			\item $\mathcal S^{(s)}_{\rm red}$ is exactly the three-term functional $\mathcal T_{\rm MS}$ in Eq.~\eqref{eq:MS_eq54} for the insulated benchmark;
			\item its Euler--Lagrange equation is the semi-implicit entropy balance
			\begin{equation}
				\delta s+\delta t\,\nabla\!\cdot
				(\mathbf j_{s,k}+\delta\mathbf j_s)
				=\sigma_k\delta t,\quad
				\delta s=\frac{c_{\rm eff}\delta T}{T_k},
				\label{eq:entropy_balance_energy_picture}
			\end{equation}
			which is the insulated form of Eq.~(57) of Ref.~\cite{Magnenet2020};
			\item under $\mathbf n\cdot\mathbf q_{k+1}=0$, the energy increment satisfies
			\begin{equation}
				\delta E_k^{\rm (3\text{-}term)}
				=\delta t\int_\Omega
				\nabla\ln T_k\cdot\delta\mathbf q\,d\omega,
				\label{eq:drift_conformal}
			\end{equation}
			and hence vanishes when $T_k$ is spatially uniform;
			\item if the source $\sigma_k$ is omitted, the Euler--Lagrange equation conserves the total entropy under insulated boundary conditions, while the energy increment is Eq.~\eqref{eq:energy_loss_2term}.
		\end{enumerate}
	\end{proposition}
	
	\begin{proof}
		Since $g^{ss}=C^{ss}=c_{\rm eff}/T_k$, the storage term is $c_{\rm eff}(\delta T)^2/(2T_k)$. Moreover,
		\[
		\left(\mathbf j_{s,k}+\frac12L\delta\mathbf F\right)\!\cdot\!\delta\mathbf F
		=-\frac1{T_k}\left(\mathbf q_k+\frac12\delta\mathbf q\right)\!\cdot\!\nabla\delta T,
		\]
		and $\sigma_k\delta T=|\mathbf q_k|^2\delta T/(\kappa T_k^2)$; these prove (i). Using $\mathbf q_{k+1}=\mathbf q_k+\delta\mathbf q$, $\mathbf q_k=-\kappa\nabla T_k$, and $\delta\mathbf q=-\kappa\nabla\delta T$, one obtains
		\[
		\nabla\!\cdot\!\left(\frac{\mathbf q_{k+1}}{T_k}\right)-\sigma_k
		=\frac1{T_k}\left[
		\nabla\!\cdot\mathbf q_{k+1}
		+\frac{\kappa}{T_k}\nabla T_k\cdot\nabla\delta T\right],
		\]
		which converts Eq.~\eqref{eq:EL_MS54_3term} into Eq.~\eqref{eq:entropy_balance_energy_picture} and proves (ii). Multiplying by $T_k$ and integrating,
		\[
		\begin{aligned}
			\int_\Omega c_{\rm eff}\delta T\,d\omega
			&=\delta t\int_\Omega\nabla\ln T_k\cdot\mathbf q_{k+1}\,d\omega\\
			&\quad+\delta t\int_\Omega T_k\sigma_k\,d\omega.
		\end{aligned}
		\]
		Since $T_k\sigma_k=-\nabla\ln T_k\cdot\mathbf q_k$, the reference-flux terms cancel, giving Eq.~\eqref{eq:drift_conformal} and proving (iii). If $\sigma_k$ is omitted, the same argument removes the production contribution and gives Eq.~\eqref{eq:energy_loss_2term}, proving (iv).
	\end{proof}
	
	\begin{remark}[Conformal-factor interpretation and the role of $\sigma_k$]
		In the energy representation, the conversion from the entropy increment to the energy increment is $\int c_{\rm eff}\delta T=\int T_k\delta s$. The local conformal factor relating the equilibrium metrics is $g^W/g^R=T_k$, and Eq.~\eqref{eq:drift_conformal} is the pairing of $\nabla\ln(g^W/g^R)$ with the heat-flux increment. In the canonical entropy representation ($\chi=-1$), the balance law generated by $\mathcal S_{\rm aug}$ is instead the internal-energy balance and conserves total energy exactly under insulated boundary conditions; its flux increment is the linearized Onsager increment $\delta\mathbf q_{\rm Ons}$. A spatially uniform reference factor, as in $\mathcal S_{\rm th}$, has zero gradient and therefore no conformal-factor drift. The source $-\sigma_k\delta T\delta t$ in Eq.~\eqref{eq:S_energy_picture} is an explicit prescribed entropy-production term added to the storage--transport sector; it is not a term generated by $\mathcal S_{\rm aug}$ itself.
	\end{remark}
	
	If the third (source) term in Eq.~\eqref{eq:MS_eq54} is omitted, the two-term ablation satisfies
	\begin{equation}
		\delta E_k^{\mathrm{(2\text{-}term)}}=-\delta t\int_\Omega
		\frac{\kappa\nabla T_k\cdot\nabla(T_k+\delta T_k)}{T_k}\,d\omega.
		\label{eq:energy_loss_2term}
	\end{equation}
	
	The strictly conservative functional with $T_{\rm ref}$ is
	\begin{equation}
		\begin{aligned}
			\mathcal S_{\rm th}[\delta T_k]
			&=\frac{1}{T_{\rm ref}^2}\int_\Omega\Bigg[
			\frac12c_{\rm eff}(\delta T_k)^2\\
			&\qquad-\left(\mathbf q_k+\frac12\delta\mathbf q_k\right)
			\cdot\nabla(\delta T_k)\,\delta t\Bigg]d\omega,
		\end{aligned}
		\label{eq:S_th_exact}
	\end{equation}
	which preserves total thermal energy identically at any finite \(\delta t\).
	
	\begin{remark}[De~Giorgi minimizing movement and distinction from Wasserstein JKO]
		\label{rem:jko}
		Let $c_{\rm eff}>0$ and $\kappa>0$ be constant, and $E_D(T)=\frac12\int_\Omega\kappa|\nabla T|^2\,d\omega$. Multiplying Eq.~\eqref{eq:S_th_exact} by $T_{\rm ref}^2/\delta t$ and dropping the additive constant $-E_D(T_k)$ gives the minimization functional
		\begin{equation}
			\frac{1}{2\delta t}\int_\Omega c_{\rm eff}|T-T_k|^2\,d\omega+E_D(T).
			\label{eq:de_giorgi_functional}
		\end{equation}
		This is the implicit-Euler minimizing-movement step for the Dirichlet-energy gradient flow in the weighted $L^2$ metric $\langle f,g\rangle_c=\int_\Omega c_{\rm eff}fg\,d\omega$~\cite{AmbrosioGigliSavare2008}. It is not the Wasserstein-metric JKO scheme originally used for the Fokker--Planck equation~\cite{JKO1998}.
	\end{remark}
	
	\paragraph{Dominant-mode estimate}
	For the dimensionless benchmark on the unit interval, take $c_{\rm eff}=1$ and $\kappa=1$, with \(T(0,x)=1+A_0\cos(\pi x)\) and \(A_0=0.6\). Then $E_0=1$. The implicit-Euler amplitude is \(A_k=A_0(1+\lambda_1\delta t)^{-k}\) with \(\lambda_1=\pi^2\). The first-mode gradient integral is
	\begin{align}
		\mathcal G(A)&=\int_0^1
		\frac{A^2\pi^2\sin^2(\pi x)}{1+A\cos(\pi x)}\,dx\nonumber\\
		&=\pi^2(1-\sqrt{1-A^2}).
		\label{eq:G_A}
	\end{align}
	The finite-time estimate is
	\begin{equation}
		\frac{\Delta E_N}{E_0}\simeq
		\frac{\kappa\lambda_1\delta t^2}{(1+\lambda_1\delta t)E_0}
		\sum_{k=0}^{N-1}\mathcal G(A_k),
	\end{equation}
	which yields $+0.73082\%$ for $\delta t=0.01$ and $+0.035479\%$ for $\delta t=5\times10^{-4}$.
	
	\section{Analytical and Numerical Verification}
	
	\subsection{Dirichlet Benchmark}
\begin{figure*}[t]
		\centering
		\includegraphics[width=0.96\textwidth]{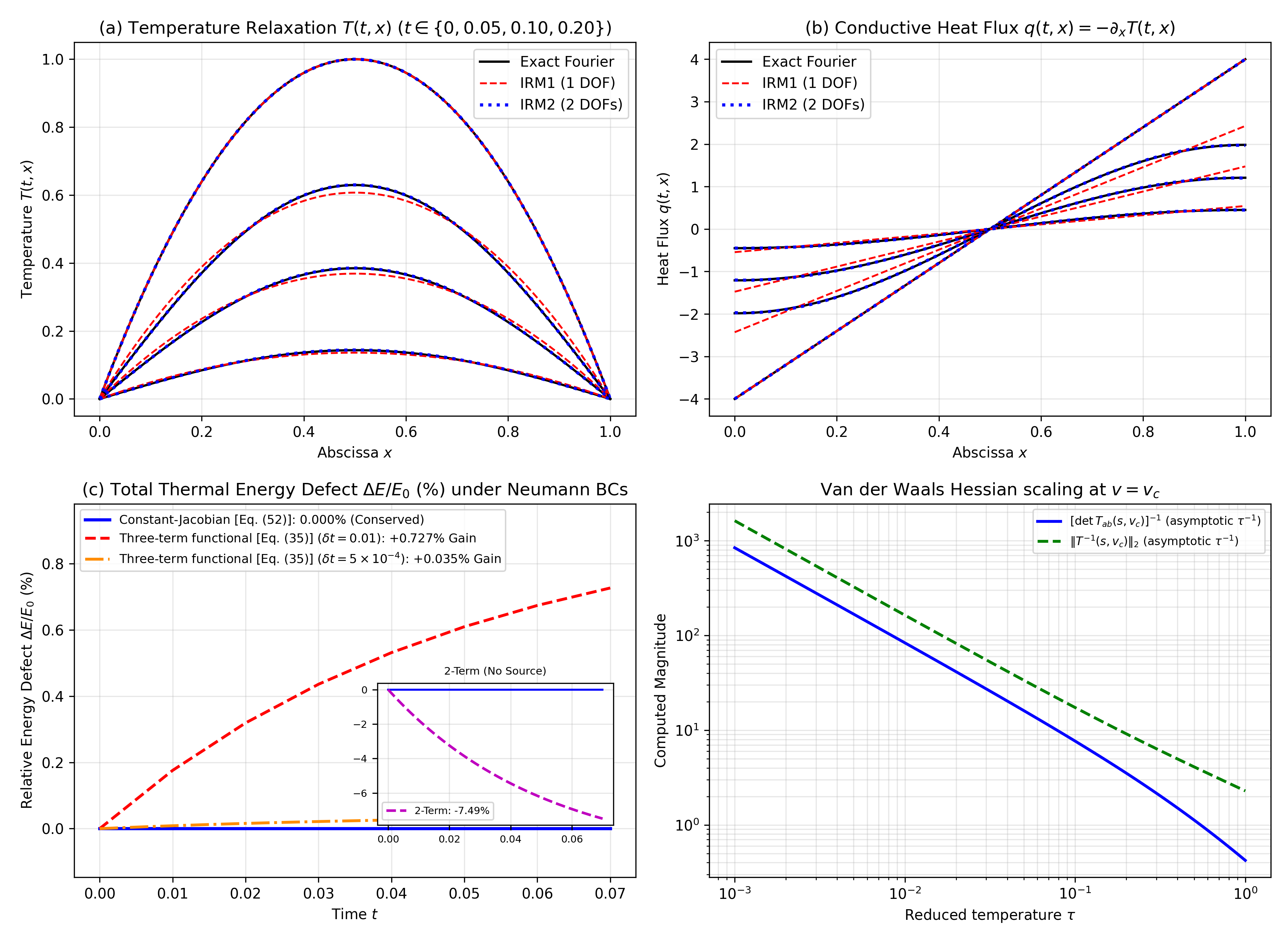}
		\caption{Numerical and analytical benchmarks. (a) Temperature relaxation. (b) Heat flux. (c) Relative energy defect. (d) Critical-isochore scaling of $[\det\mathbf T]^{-1}$ (blue solid) and $\|\mathbf T^{-1}\|_2$ (green dashed), evaluated in the fixed $(s,v)$ coordinate chart.}
		\label{fig:verification}
	\end{figure*}

	We consider the dimensionless heat equation
	\[
	\begin{aligned}
		\partial_t T&=\partial_{xx}T,\qquad 0<x<1,\\
		T(t,0)&=T(t,1)=0.
	\end{aligned}
	\]
	with $T(0,x)=4x(1-x)$. Its exact Fourier solution is
	\[
	T_{\rm ex}(t,x)=\sum_{\substack{n\geq1\\ n\,\mathrm{odd}}}
	\frac{32}{\pi^3n^3}e^{-n^2\pi^2t}\sin(n\pi x).
	\]
	For a trial basis $\{N_i\}$, the backward-Euler Ritz update is
	\[
	\begin{aligned}
		(\mathbf M+\delta t\,\mathbf K)\mathbf a_{k+1}
		&=\mathbf M\mathbf a_k,\\
		M_{ij}&=\int_0^1N_iN_j\,dx,\\
		K_{ij}&=\int_0^1N_i'N_j'\,dx.
	\end{aligned}
	\]
	For IRM1 with $N_1=x(1-x)$, the one-dimensional mass and stiffness
	matrices are $M_1=1/30$ and $K_1=1/3$. Thus the Rayleigh quotient is
	$K_1/M_1=10$, and minimization gives
	$A_{k+1}=A_k/(1+10\delta t)$, i.e.\
	$\lambda_1^{\rm IRM1}=10$ (relative error $1.32\%$).
	For IRM2, with basis $\{x(1-x),x^2(1-x)^2\}$, Beta-function integration gives
	\[
	\mathbf M=\begin{pmatrix}\tfrac1{30}&\tfrac1{140}\\
		\tfrac1{140}&\tfrac1{630}\end{pmatrix},\qquad
	\mathbf K=\begin{pmatrix}\tfrac13&\tfrac1{15}\\
		\tfrac1{15}&\tfrac2{105}\end{pmatrix}.
	\]
	The generalized eigenvalue equation is
	$\det(\mathbf K-\lambda\mathbf M)=0$, or
	$\lambda^2-112\lambda+1008=0$. By the Courant--Fischer
	characterization,
	\[
	\lambda_1^{\rm IRM2}=56-4\sqrt{133}\approx9.8697496>\pi^2
	\]
	(relative error $1.5\times10^{-5}$). Both trial modes are symmetric about
	$x=1/2$, so the second Ritz value
	$56+4\sqrt{133}\approx102.13$ must be compared with $9\pi^2$,
	not $4\pi^2$. Results are shown in Fig.~\ref{fig:verification}(a,b)
	and Table~\ref{tab:summary}.

	\subsection{Neumann Energy-Conservation Benchmark}
	For $T_0(x)=1+0.6\cos(\pi x)$ on $[0,1]$, one has $T_-=0.4$, $G_0=0.6\pi/\sqrt{2}$, and $H_0=0.6\pi^2/\sqrt{2}$. Taking $C_L=2$ and the $3/4$ absorption threshold gives the sufficient restriction $\delta t\leq\delta t_0\approx1.07\times10^{-3}$. Thus the finer tested step $5\times10^{-4}$ lies inside the sufficient time-step range of the continuum one-dimensional theorem, whereas $0.01$ lies outside it. The sufficient restriction is conservative: the proof uses the absorption inequality at every step, whereas $\|T_k''\|_{L^2}$ decays rapidly in practice. The lumped-mass $P_1$ run at $\delta t=0.01$ lies outside the sufficient range but remains stable and convergent in all reported tests. Proving a sharper threshold, or establishing unconditional stability of the lumped-mass $P_1$ implementation, is left among the open problems.
	
	We use \(M=101\) nodes, lumped mass, and element-midpoint coefficients on \(x\in[0,1]\). The lumped-mass implementation gives: constant-Jacobian $|\Delta E/E_0| \le 2.2 \times 10^{-15}$; three-term $+0.7269\%$ and $+0.03546\%$; two-term $-7.4897\%$. The ratios $0.73082/0.035479\approx 20.6$ and $0.7269/0.03546\approx 20.5$ are close to the ratio $20$ of the time steps. Since the number of steps scales as $\delta t^{-1}$, this corresponds to $\ln 20.6/\ln 20\approx 1.00$, consistent with the per-step $O(\delta t^2)$ drift of Proposition~\ref{prop:drift_bound}.

	\subsection{Mixed \texorpdfstring{$DG_0$--$RT_0$}{DG0--RT0} benchmark with fully implicit nonlinear conductivity}
	\label{sec:dg_rt_benchmark}
	The dimensionless model with \(\kappa(T)=1+T^2\) is solved on $x\in[0,1]$ with $T(0,x)=1+0.6\cos(\pi x)$, $c_{\rm eff}=1$, $\delta t=5\times10^{-4}$, $t_{\rm final}=0.07$, insulated boundary conditions, and a damped Newton iteration with residual tolerance $10^{-12}$. The cell-energy update $\delta e_i=-\delta t\,[q_h(x_{i+1})-q_h(x_i)]/|K_i|$ is exact by construction, so the energy defect is at roundoff ($\approx10^{-14}\%$), independent of the Newton tolerance.
	
	\begin{proposition}[Discrete conservation of the $DG_0$--$RT_0$ scheme]
		\label{prop:dg_rt_conservation}
		Let $\delta e_h\in DG_0(\Omega_h)$ and $q_h\in RT_0^0(\Omega_h)$ be the discrete energy and flux increments at step $k+1$. If the discrete balance equation $|K_i|\,\delta e_i+\delta t\,[q_h(x_{i+1})-q_h(x_i)]=0$ holds on every cell with $q_h(x_0)=q_h(x_N)=0$, then $E_{k+1,h}-E_{k,h}=-\delta t\,[q_h(x_N)-q_h(x_0)]=0$, independently of the conductivity law $\kappa(T)$.
	\end{proposition}
	\begin{proof}
		Summing the cellwise balance equations over $i$ telescopes the flux-difference sum, leaving only the boundary flux, which vanishes by hypothesis.
	\end{proof}
	
	\begin{remark}[Numerical consistency check of Proposition~\ref{prop:exact_elimination}]
		\label{rem:prop3_verification}
		We solved the reduced weak equation~\eqref{eq:reduced_weak_galerkin} on the CG1 discretization with $N\in\{20,40,80,160\}$ elements, $T_k(x)=1+0.1\cos(\pi x)$, $c_{\rm eff}=\kappa=1$, and $\delta t=10^{-3}$. The reduced solution $\delta I_h$ is non-zero, smooth, and converged at $N=160$ ($\|\delta I_h\|_\infty\approx1.2\times10^{-3}$; the values at $N=20$ and $N=160$ agree to the fourth digit). The algebraic reconstruction of Eq.~\eqref{eq:exact_reconstruction} then gives $\delta e_h$, $\delta j_h$, and $\delta\phi_h$ with $\lambda_h=\mu_h=0$, and the five algebraic Euler--Lagrange blocks vanish identically in IEEE double precision ($\max|r_\phi|=\max|r_e|=\max|r_j|=\lambda_h=\mu_h=0.0$ exactly), while the weak residual is at $\mathcal{O}(10^{-20})$.
		
		Because the reconstruction relations~\eqref{eq:exact_reconstruction} directly enforce the zero-residual conditions, the vanishing of the unperturbed algebraic blocks verifies machine-precision consistency of the elimination map. To test the sensitivity of the residual operators away from the constraint set, we perturbed $\delta e_h$ and $\delta j_h$ by a factor $(1+\epsilon)$ with $\epsilon=10^{-6}$. The residuals then scale linearly as $|r_\phi|\approx1.08\times10^{-9}$, $|r_e|\approx9.79\times10^{-10}$, and $|r_j|\approx3.13\times10^{-9}$ at $N=160$, matching the expected $\epsilon\|\delta e_h\|_\infty\approx9.8\times10^{-10}$, $\epsilon\|I_k\delta e_h\|_\infty\approx1.08\times10^{-9}$, and $\epsilon\|\delta j_h\|_\infty\approx3.13\times10^{-9}$. The ratio $|r_j|/|r_e|\approx3.2$ is consistent with $\|\delta j_h\|_\infty/\|\delta e_h\|_\infty\approx3.2$. The residuals scale linearly with $\epsilon$, as they must for the algebraic residual operators; this verifies the implementation of the residuals and is not independent evidence for the elimination identity, which is an algebraic statement. An independent saddle-point solve of the full augmented system, treating all six blocks as independent unknowns, remains an open problem.
	\end{remark}
	
	Table~\ref{tab:mixed_nonlinear} reports the raw cell-centred differences $d_N$ against the $N=320$ reference and the observed convergence orders $p_N=\log_2(d_{N/2}/d_N)$. Because the reference is itself a numerical solution rather than an exact solution, these values are descriptive and are not an independent proof of an asymptotic convergence order.

	\begin{table*}[t]
		\centering
		\caption{Fully implicit nonlinear mixed benchmark with $\kappa(T)=1+T^2$. The quantity $d_N$ is the cell-centred discrete $L^2$ difference from the $N=320$ cell-average reference. The reported $p_N=\log_2(d_{N/2}/d_N)$ is the observed convergence order, not the raw error ratio. For an exactly second-order error $e_N=Ch_N^2$ one has $d_N=C(h_N^2-h_{\rm ref}^2)$ with $h_{\rm ref}=1/320$, which predicts $p_N=2.017,\,2.070,\,2.322$ for $N=40,80,160$ ($C\approx0.1137$); these predictions agree with the tabulated values to within about $2.0\times10^{-3}$. These values are descriptive; the growth of $p_N$ is an artefact of the nearby reference, not a loss of accuracy.}
		\label{tab:mixed_nonlinear}
		\small
		\begin{tabular}{rrrrrr}
			\toprule
			$N$ & $h$ & $d_N$ & observed order $p_N$ & max $|\Delta E/E_0|$ (\%) & max residual\\
			\midrule
			20  & $5.0\times10^{-2}$ & $2.82736\times10^{-4}$ & --    & $5.55\times10^{-14}$ & $9.41\times10^{-13}$\\
			40  & $2.5\times10^{-2}$ & $6.99476\times10^{-5}$ & 2.015 & $3.33\times10^{-14}$ & $8.68\times10^{-13}$\\
			80  & $1.25\times10^{-2}$ & $1.66599\times10^{-5}$ & 2.070 & $2.22\times10^{-14}$ & $9.66\times10^{-13}$\\
			160 & $6.25\times10^{-3}$ & $3.33225\times10^{-6}$ & 2.322 & $4.44\times10^{-14}$ & $8.71\times10^{-13}$\\
			\bottomrule
		\end{tabular}
	\end{table*}
	
	\begin{figure*}[t]
		\centering
		\includegraphics[width=0.96\textwidth]{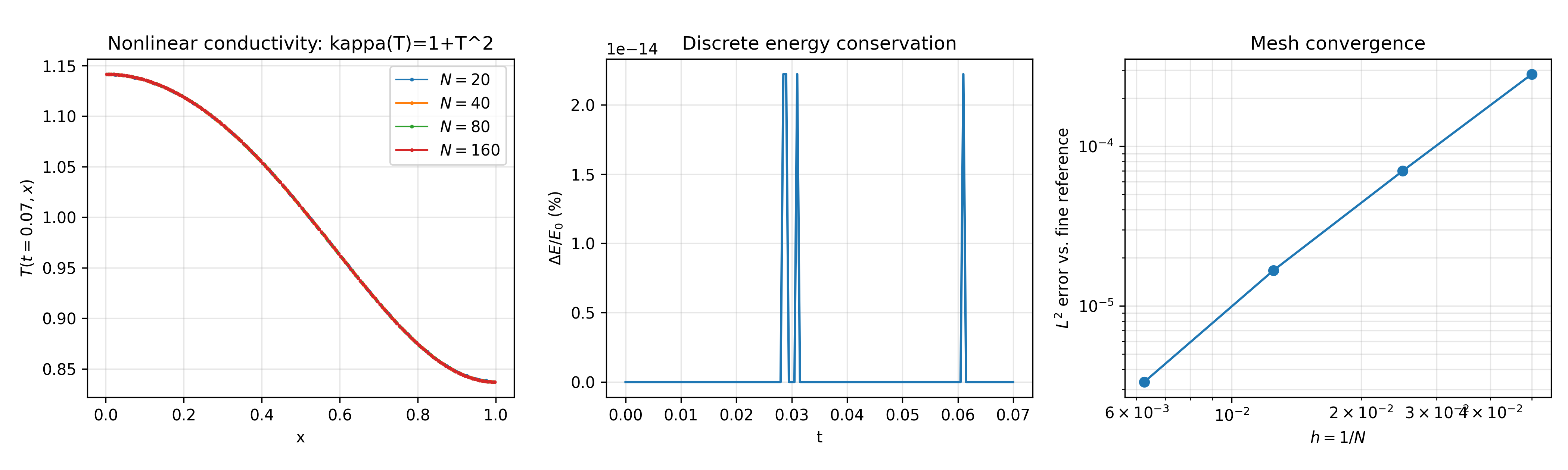}
		\caption{Fully implicit mixed $DG_0$--$RT_0$ benchmark for $\kappa(T)=1+T^2$.}
		\label{fig:mixed_nonlinear}
	\end{figure*}
	
	The Python scripts used to generate Figures~\ref{fig:verification} and \ref{fig:mixed_nonlinear}, together with the numerical summary for the mixed $DG_0$--$RT_0$ benchmark, are provided as Supplementary Material S1.
	
	\begin{table*}[t]
		\centering
		\caption{Summary of analytical and numerical benchmarks. Energy defects are outputs of the lumped-mass $P_1$ implementation; the van der Waals exponent is asymptotic ($\tau\to0^+$).}
		\label{tab:summary}
		\small
		\begin{tabular}{p{3.6cm}p{5.6cm}p{7.2cm}}
			\toprule
			Benchmark & Analytical prediction & Numerical result\\
			\midrule
			First Dirichlet mode & $\pi^2\approx9.869604$ & $10$ (IRM1); $56-4\sqrt{133}\approx9.8697496$ (IRM2)\\
			Second symmetric mode & $9\pi^2\approx88.82644$ & $56+4\sqrt{133}\approx102.13025$ (overestimate $\approx14.98\%$)\\
			Neumann defect at $t=0.07$ & first-mode: $+0.73082\%$, $+0.035479\%$ ($\delta t=0.01,\,5\times10^{-4}$); constant Jacobian: $0$ & lumped $P_1$: $+0.7269\%$, $+0.03546\%$; two-term $-7.4897\%$; constant Jacobian $\le2.2\times10^{-15}$\\
			Van der Waals isochore & $[\det T]^{-1}\propto\tau^{-1}$ & exact isochore expression, Eq.~\eqref{eq:vdw_isochore}\\
			\bottomrule
		\end{tabular}
	\end{table*}
	
	\section{Van der Waals Critical-Isochore Scaling}
	Consider the van der Waals fluid in the energy representation ($\chi=+1$),
	\begin{equation}
		\begin{aligned}
			u(s,v)&=(v-b)^{-2/3}e^{2s/3}-\frac{a}{v},\\
			T&=\frac{2}{3}(v-b)^{-2/3}e^{2s/3}.
		\end{aligned}
		\label{eq:vdw_u}
	\end{equation}
	The exact Tisza Hessian in coordinates $(s,v)$ is
	\begin{equation}
		\mathbf T=\begin{pmatrix}
			\dfrac{2T}{3} & -\dfrac{2T}{3(v-b)}\\[5pt]
			-\dfrac{2T}{3(v-b)} & \dfrac{5T}{3(v-b)^2}-\dfrac{2a}{v^3}
		\end{pmatrix},
		\label{eq:vdw_hessian}
	\end{equation}
	with determinant
	\begin{equation}
		\det\mathbf T=\frac{2T}{3v^3(v-b)^2}\left[Tv^3-2a(v-b)^2\right].
		\label{eq:detT_vdw}
	\end{equation}
	The critical point is $v_c=3b$ and $T_c=8a/(27b)$. On the critical isochore, set $T=T_c(1+\tau)$ with $\tau\to0^+$. Direct substitution gives
	\begin{equation}
		\det\mathbf T(s,v_c)=\frac{32a^2}{2187b^4}(1+\tau)\tau.
		\label{eq:vdw_isochore}
	\end{equation}
	Hence $[\det\mathbf T]^{-1}\propto\tau^{-1}$ on the critical isochore, and the equilibrium Hessian becomes singular as $\tau\to0^+$.
	
	In the fixed $(s,v)$ coordinate chart, the spectral norm of the inverse
	Hessian has the same leading critical exponent. Let
	$\mathbf T_{\rm crit}:=\lim_{\tau\to0^+}\mathbf T(s(\tau),v_c)$;
	its trace is finite and positive. For $\tau>0$, $\mathbf T$ is symmetric
	positive definite and
	$\lambda_{\max}(\mathbf T)\to\operatorname{tr}\mathbf T_{\rm crit}$.
	Since $\det\mathbf T=\lambda_{\min}(\mathbf T)\lambda_{\max}(\mathbf T)$,
	Eq.~\eqref{eq:vdw_isochore} gives
	\begin{equation}
		\begin{aligned}
			\lambda_{\min}(\mathbf T)
			&=\frac{\det\mathbf T}{\lambda_{\max}(\mathbf T)}
			\propto\tau,\\
			\|\mathbf T^{-1}(s,v_c)\|_2
			&=\lambda_{\min}(\mathbf T)^{-1}\propto\tau^{-1}.
		\end{aligned}
	\end{equation}
	This is the green dashed curve in Fig.~\ref{fig:verification}(d); the spectral norm is a coordinate-dependent diagnostic in the stated fixed chart.
	
	For the reduced, linearized conserved-density diffusion model, assume the mobility matrix obeys uniform spectral bounds
	$0<L_-\mathbf I\preceq\mathbf L(\tau)\preceq L_+\mathbf I$
	with $L_-$ and $L_+$ independent of $\tau$. Then, provided the soft
	Hessian eigendirection belongs to the conserved sector, the slowest
	nonzero diffusive rate at fixed nonzero wave vector $\mathbf k$ satisfies
	$\Gamma_{\min}(\mathbf k,\tau)\asymp |\mathbf k|^2\tau$.
	This is a conditional Model-B-type critical-slowing estimate, not a
	classification of the full hydrodynamic critical dynamics of a real fluid~\cite{Hohenberg1977}.

	\section{Concluding Remarks}
	We have formulated a constrained mixed variational principle with Legendre-dual defects. The quadratic equilibrium defect is a Fenchel--Young gap, and its nonlinear Fenchel--Young extension is self-dual under the stated partial Legendre transformation; the dissipative defect is a separate Fenchel--Young gap for a general convex dissipation potential (Proposition~\ref{prop:nonlinear_dissipation}). For the natural nonlinear storage--defect block, the storage mismatch in Eq.~\eqref{eq:nonlinear_storage_mismatch} vanishes for centered degree-two homogeneous potentials but not in general. The exact elimination result of Proposition~\ref{prop:exact_elimination} shows that the multiplier and contact fields are reconstructible algebraically from the reduced weak balance problem, so no independent saddle-point solve is required. Together with the weighted drift identity (Proposition~\ref{prop:drift_bound}) and the one-dimensional a priori stability theorem (Proposition~\ref{prop:1d_weighted_stability}), these results provide a self-contained variational and stability analysis of the incremental heat-conduction benchmark.
	
	An extended-fibre structure related to Quevedo geometry, the Hessian transport defect for genuinely transport-dependent potentials, and the curvature of a potential-weighted Hessian metric distinct from the standard Quevedo $G^{II}$ pullback are developed in a companion paper.
	
	Open problems and possible extensions include: (i) constructing a fully nonlinear augmented action that preserves the physical balance law and is invariant off shell under the full Legendre transformation; (ii) implementing the exact elimination of Proposition~\ref{prop:exact_elimination} in a fully compatible Galerkin code for the full augmented system and reporting all reconstructed residuals; (iii) extending the weighted drift identity beyond constant $\kappa$; (iv) extending the weighted-scheme a priori stability theorem from one space dimension to dimensions two and three and proving mesh-uniform bounds for a fully discrete Galerkin realization; (v) continuation toward singular Hessian boundaries $\det T=0$, which requires additional regularization; and (vi) comparison with path-dependent formulations on a common benchmark.

	\section*{Acknowledgments and Availability}
	The author used an AI assistant during the preparation of this manuscript to review mathematical derivations and algebraic identities. The scientific contributions, numerical implementations, physical interpretations and conclusions are the author's own, and the author takes full responsibility for the accuracy and integrity of the manuscript. The Python scripts used to generate Figures~\ref{fig:verification} and \ref{fig:mixed_nonlinear}, together with the numerical summary for the mixed finite-element benchmark, are supplied as Supplementary Material S1. A public repository with a persistent DOI may be added in a future version.
	
	\appendix
	\titleformat{\section}{\large\bfseries}{Appendix \Alph{section}.}{0.5em}{}
	
	\section{Proof of Proposition~\ref{prop:1d_weighted_stability}}
	\label{app:stability_1d}
	
	Write $b_k=(\ln T_k)'$ and $B_k=\|b_k\|_{L^\infty}$. Testing Eq.~\eqref{eq:1d_weighted_step} by $-T_{k+1}''$, integrating by parts, and using the Neumann conditions gives
	\begin{equation}
		\begin{aligned}
			\|T_{k+1}'\|_2^2-\|T_k'\|_2^2+\|v_k'\|_2^2
			&+\frac{\kappa\delta t}{c}\|T_{k+1}''\|_2^2\\
			&\leq\frac{\kappa\delta t}{c}B_k^2\|v_k'\|_2^2.
		\end{aligned}
		\label{eq:1d_energy_pre_app}
	\end{equation}
	This follows from $\int_0^L v_k'T_{k+1}'\,dx=\frac12(\|T_{k+1}'\|_2^2-\|T_k'\|_2^2+\|v_k'\|_2^2)$ and Young's inequality. Hence, whenever $\alpha_k:=(\kappa\delta t/c)B_k^2\leq3/4$,
	\begin{equation}
		\|T_{k+1}'\|_2^2-\|T_k'\|_2^2+\frac14\|v_k'\|_2^2+\frac{\kappa\delta t}{c}\|T_{k+1}''\|_2^2\leq0.
		\label{eq:1d_energy_absorbed_app}
	\end{equation}
	
	The one-dimensional interpolation inequalities are
	\begin{equation}
		\begin{aligned}
			\|f\|_\infty^2&\leq C_L\bigl(\|f\|_2^2+\|f\|_2\|f'\|_2\bigr),\\
			\|f\|_4^2&\leq C_L\bigl(\|f\|_2^2+\|f\|_2^{3/2}\|f'\|_2^{1/2}\bigr).
		\end{aligned}
	\end{equation}
	They hold with a common constant depending only on $L$; on $[0,1]$, $C_L=2$ is admissible. Applying the first inequality to $f=T_k'$ yields
	\begin{equation}
		B_k^2\leq\frac{C_L}{T_-^2}\left(\|T_k'\|_2^2+\|T_k'\|_2\|T_k''\|_2\right).
		\label{eq:1d_B_bound}
	\end{equation}
	If $G_0=0$, then $T_0$ is constant; this constant profile solves every time step, and uniqueness follows from the coercivity argument below. All estimates in the proposition are then immediate. Hence assume $G_0>0$ in what follows. Set
	\begin{align*}
		\delta t_{\rm init}
		&:=\frac{3cT_-^2}{4\kappa C_L(G_0^2+G_0H_0)},\\
		s_*&:=\frac{\sqrt{1+3T_-^2/(C_LG_0^2)}-1}{2}.
	\end{align*}
	Define additionally
	\[
	\delta t_{\rm rec}:=\frac{cs_*^2}{\kappa},
	\qquad
	\delta t_0:=\min\{\delta t_{\rm init},\delta t_{\rm rec}\}.
	\]
	At $k=0$, Eq.~\eqref{eq:1d_B_bound} gives $\alpha_0\leq3/4$. Suppose inductively that Eq.~\eqref{eq:1d_energy_absorbed_app} holds through step $k-1$. Then $\|T_k'\|_2\leq G_0$ and $\|T_k''\|_2\leq G_0\sqrt{c/(\kappa\delta t)}$. With $s=\sqrt{\kappa\delta t/c}$,
	\[
	\alpha_k\leq\frac{C_LG_0^2}{T_-^2}(s^2+s)\leq\frac34,
	\]
	where the last inequality is precisely $\delta t\leq\delta t_{\rm rec}$. This closes the induction and proves Eqs.~\eqref{eq:1d_H1_uniform} and \eqref{eq:1d_L2H2}.
	
	For stepwise solvability, the weak equation for $v_k$, after testing by $w/T_k$, has the bilinear form $a_k(v,w)=\frac{c}{\delta t}\int_0^L\frac{vw}{T_k}\,dx+\kappa\int_0^L\frac{v'w'}{T_k}\,dx$, coercive in $H^1(0,L)$. Lax--Milgram gives a unique weak solution; the equation and $b_k\in L^\infty(0,L)$ then give $v_k''\in L^2(0,L)$, hence $T_{k+1}\in H^2(0,L)$.
	
	For temperature bounds, rewrite the update as $(\frac{c}{\delta t}-\kappa\partial_{xx}+\kappa b_k\partial_x)T_{k+1}=\frac{c}{\delta t}T_k+\kappa|T_k'|^2/T_k$. The comparison principle with the constant subsolution $T_-$ gives $T_{k+1}\geq T_-$. The constant supersolution $\|T_k\|_\infty+(\kappa\delta t/(cT_-))\|T_k'\|_\infty^2$ gives $\|T_{k+1}\|_\infty\leq\|T_k\|_\infty+\frac{\kappa\delta t}{cT_-}\|T_k'\|_\infty^2$.
	
	Set $a_k=\|T_k''\|_2$. The energy estimate gives $\sum_{k=0}^{N-1}\delta t\,a_k^2\leq C_2$ with $C_2:=\delta t_0H_0^2+(c/\kappa)G_0^2$. Applying the first interpolation inequality to $T_k'$ and summing in time yields
	$\sum_{k=0}^{N-1}\delta t\|T_k'\|_\infty^2\leq C_L(T_fG_0^2+G_0\sqrt{T_fC_2})$.
	Consequently one may take $M_T=\|T_0\|_\infty+(\kappa C_L/(cT_-))(T_fG_0^2+G_0\sqrt{T_fC_2})$.
	
	Finally, the second interpolation inequality applied to $f=T_k'$ gives $\|R_k/T_k\|_2\leq(\kappa/T_-)a_k$ and $\|\vartheta_k\|_2\leq(\kappa/T_-)a_k+(\kappa C_L/T_-^2)(G_0^2+G_0^{3/2}a_k^{1/2})$. Hence
	\[
	\sum_{k=0}^{N-1}\delta t\|\vartheta_k\|_2\|R_k/T_k\|_2\leq C_P,
	\]
	Here the constant $C_P$ is defined by
	\[
	\begin{aligned}
		C_P:=&\frac{\kappa^2C_2}{T_-^2}
		+\frac{\kappa^2C_LG_0^2}{T_-^3}\sqrt{T_fC_2}\\
		&+\frac{\kappa^2C_LG_0^{3/2}}{T_-^3}T_f^{1/4}C_2^{3/4}.
	\end{aligned}
	\]
	The bound follows from Cauchy--Schwarz and H\"older's inequality in time. Proposition~\ref{prop:drift_bound}, with $T_k\leq M_T$ and $m_0=c/M_T$, gives Eq.~\eqref{eq:1d_cumulative_drift}. All constants depend on $T_f$, initial data, $L$, $T_-$, $c$, and $\kappa$, but not on $\delta t$ or $N$.

	\end{document}